\documentclass[epj,epsfig,graphics]{svjour}
\usepackage{latexsym}
\usepackage{graphicx}
\begin{document}

\title{Scalable Ion Trap Quantum Computing without Moving Ions}
\author{L. Tian\inst{1,3} \and R. Blatt\inst{2,3} \and P. Zoller\inst{1,3}}
\institute{Institute for Theoretical Physics, University of Innsbruck, 6020 Innsbruck,
Austria \and Institute for Experimental Physics, University of Innsbruck,
6020 Innsbruck, Austria \and Institute for Quantum Optics and Quantum
Information of the Austrian Academy of Sciences, 6020 Innsbruck, Austria}
\date{Received: date / Revised version: date}

\abstract{A hybrid quantum computing scheme is studied where the
hybrid qubit is made of an ion trap qubit serving as the
information storage and a solid-state charge qubit serving as the
quantum processor, connected by a superconducting cavity. In this
paper, we extend our previous work\cite{interfacing} and study the
decoherence, coupling and scalability of the hybrid system. We
present our calculations of the decoherence of the coupled ion -
charge system due to the charge fluctuations in the solid-state
system and the dissipation of the superconducting cavity under
laser radiation. A gate scheme that exploits rapid state flips of
the charge qubit to reduce decoherence by the charge noise is
designed. We also study a superconducting switch that is inserted
between the cavity and the charge qubit and provides tunable
coupling between the qubits. The scalability of the hybrid scheme
is discussed together with several potential experimental
obstacles in realizing this scheme.
}

\maketitle

\section{Introduction}

\label{sec:1} Ion trap quantum computing has achieved great
progresses in the past few years. On the experimental side,
controlled quantum logic gate and quantum teleportation have been
demonstrated\cite{ion_trap_exp}; on the theory side, scalable
schemes by moving the ions\cite{Wineland2002} and fast quantum
logic gates have been proposed\cite{PhysToday2004}. One impending
question at the moment is to build scalable ion trap quantum
computing systems that can perform quantum algorithms beyond the
simple demonstration level. In a previous
publication\cite{interfacing}, we studied a scalable quantum
computing scheme that connects a quantum optical qubit and a
solid-state qubit into a hybrid qubit\cite{interfacing}. Quantum
optical qubits have long life time; and solid-state qubits can
perform fast quantum logic gates on a nanosecond time scale. By
interfacing the two systems, we hope to combine the best of the
two systems, given that the two systems are compatible with each
other. One example system is the ion trap qubit connecting with
the superconducting charge qubit
\cite{superconducting_qubits,loss_divincenzo_dots,divincenzo_exp_quant_comp_2000}.
The ion qubit, made of the internal mode of the ion, bears the
tasks of single qubit gate and information storage. The
superconducting qubit bears the tasks of controlled gates, qubit
detection and quantum state transport.

A key question in this scheme is the coupling between the quantum
optical and solid-state qubit, allowing the swap of the states of
the two qubits. By applying a polarization dependent laser pulse,
the internal mode of the ion is coupled with the motional mode of
the ion; the charge qubit couples with the motional mode via
capacitive coupling\cite{cpb_mechanical_resonator_schwab}. Hence
the motion is an effective connection between the two
qubits\cite{heinzen_wineland_trap_couping_1990}. Exchange of
information is achieved through a swap gate between the two
qubits. We showed that a fast swap gate that is independent of the
motional state can be achieved. A superconducting cavity is
inserted between the ion and the charge qubit to: 1. increase the
magnitude of the coupling; 2. ensure compatibility -- to prevent
the stray photons of the ion trap from radiating the charge qubit.

In this paper, we extend our previous work on the hybrid qubit
scheme and study several practical issues in the implementation of
this scheme. We study the decoherence of the coupled qubits due to
various environmental noise, such as the charge fluctuations in
the solid-state system and the dissipation of the superconducting
cavity under the stray photons. We show that by exploiting a rapid
state flipping technique during the swap gate, the effect of the
charge fluctuations can be largely reduced. We also discuss issues
concerning compatibility and scalability when combining very
different systems together. The paper is organized as follows. In
section 2, we review the protocol of the hybrid qubit quantum
computing, the Hamiltonian of the coupled system and the fast
quantum phase gate between the qubits. In section 3, we study
the decoherence due to various noise and present a gate scheme
that reduces the effect of the low frequency ($1/f$) charge noise.
In section 4, we discuss several experimental issues, including
the fast switch of the capacitive coupling, the decoupling of the
charge qubit from the ac driving of the trap, and the scalability
of the scheme. Finally, in section 5, we discuss potential
technical obstacles in realizing this scheme and conclude this
paper.

\section{The System}

\label{sec:2} In this section, we review the concept of hybrid
qubit quantum computing and an implementation of the hybrid qubit
with a trapped ion qubit and a superconducting charge qubit
\cite{interfacing}. The strength of the coupling between the ion
qubit and the charge qubit is increased by inserting a
superconducting cavity between the qubits.

\subsection{The Hybrid Qubit Scheme}

\label{sec:2.1}
\begin{figure}[tbp]
\begin{center}
\includegraphics[width=5cm]{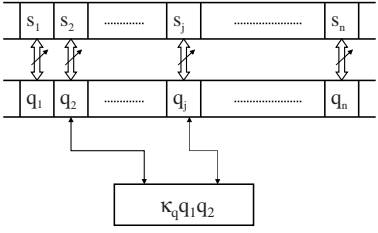}
\end{center}
\caption{The operating protocol of the composite qubits. For
explanation see text.} \label{fig:1}
\end{figure}

The scheme is shown in Fig.~\ref{fig:1} where the
$\left\{ s_{i}\right\} $ blocks are the quantum optical qubits
which serve as the storage elements and the $\left\{
q_{i}\right\}$ blocks are the solid-state qubits which play the
role of the processing elements. The state of the system during
storage is $|\{q_{i}=0\} \rangle \sum c_{\{s_{i}\}}|\{
s_{i}\}\rangle$ where $\{s_{i}\}$ ($\{q_{i}\}$) includes all
quantum optical (solid-state charge) qubits. Here the charge
qubits are in their ground state and the quantum optical qubits
are decoupled from the solid-state qubit. Initialization of the
quantum optical qubits can be achieved by optical pumping. A two
bit gate between $s_{i}$ and $s_{j}$ can be achieved via the
charge qubits. First, the swap gates between qubits $\left\{
s_{i},q_{i}\right\}$ and $\left\{ s_{j},q_{j}\right\}$ are applied
to give the state $|\{ q_{i}^{\prime}=0 \}\rangle \sum
c_{\{s_{i}^{\prime}\}}|\{s_{i}^{\prime}\} \rangle$, where
$\{s_{i}^{\prime}\}$($\{q_{i}^{\prime}\}$) includes the spin
(charge) qubits $s_{k}$ ($q_{k}$) with $k\ne\, i,j$ and the two
charge (spin) qubits $q_{i}$ and $q_{j}$ ($s_{i}$ and $s_{j}$).
Then the two bit gate is applied on the charge qubits
$q_{i},q_{j}$ and gives $|\{ q_{i}^{\prime}=0 \}\rangle \sum
\bar{c}_{\{s_{i}^{\prime}\}}|\{s_{i}^{\prime}\} \rangle$ with
coefficient $\bar{c}_{\{s_{i}^{\prime}\}}$ different from
$c_{\{s_{i}^{\prime}\}}$. Finally, the swap gates transfer the
states in $q_{i},q_{j}$ back to $s_{i},s_{j}$ with the state $|\{
q_{i}=0 \}\rangle \sum \bar{c}_{\{s_{i}\}}|\{s_{i}\} \rangle$, and
a two bit gate between the spin qubits $s_{i}$ and $s_{j}$ is
achieved.

\begin{figure}[hbt]
\begin{center}
\includegraphics[width=7.5cm]{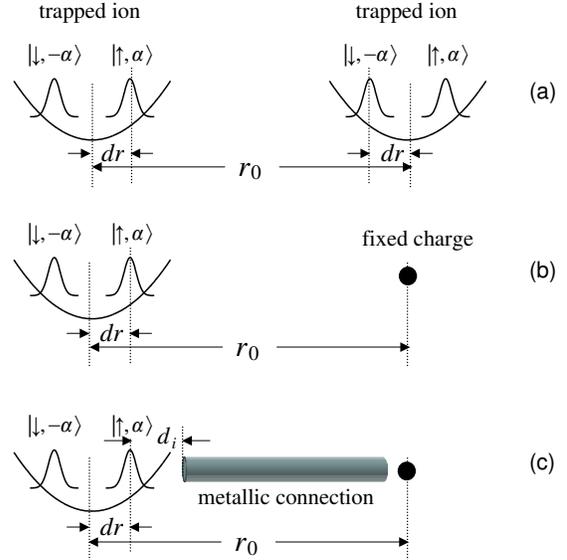}
\end{center}
\caption{The interactions: (a) dipole-dipole interaction between
two ions with distance $r_{0}$ and laser induced spatial
displacement $\pm d_r$ for the internal states $|\uparrow\rangle$
and $| \downarrow\rangle$ respectively; (b) dipole-charge
interaction between an ion and a fixed charge;  (c) dipole-charge
interaction mediated by a metallic connection with a distance
$d_{i}$ between the ion and the connection.} \label{fig:2}
\end{figure}

The generic Hamiltonian of the combined system can be written as
$H_{t}=H_{s}+H_{q}+H_{\mathrm{int}}$. Here
$H_{s}$\cite{ion_trap_decoherence_review} describes an
harmonically trapped ion manipulated by laser pulses with the
motional energy $\hbar \omega _{\nu } \hat{a}^{\dag }\hat{a}$,
laser detuning $\delta _{0}\sigma _{z}^{s}$ and the Rabi flipping
term $\hbar \omega _{R}\left( \sigma _{+}^{s}e^{i{_{l}\delta
k_{l}\hat{x}}}+\mathrm{h.c.}\right)$. Here, $\hat{x}$ is the
coordinate of the ion, $\hat{a}$($\hat{a}^\dagger$) is the
lowering(raising) operator, and $\omega _{\nu }$ the trapping
frequency. The term $H_{q}=\frac{E_{z}}{2}\sigma
_{z}^{q}+\frac{E_{x}}{2}\sigma _{x}^{q}$ describes a solid-state
qubit with the generic form of a quantum two level system with
energies $E_{x,z}$. The coupling $H_{\mathrm{int}}=\hbar \kappa
(t)\hat{x}\sigma _{z}^{q}$ describes a fixed charged particle
interacting with a harmonically trapped charged particle.  Note
the laser pulse generates coupling between the motion $\hat{x}$ of
the ion and the internal mode $\sigma _{z}^{s}$ of the ion; hence
generates an indirect coupling between the internal mode and the
solid-state qubit. The coupling amplitude is of the order of
$(\sigma_{z}^{s}\sigma_{z}^{q}) Qe dr/4\pi \epsilon_{0}r_{0}^{2}$
describing the interaction between a charge $Q$ and a dipole $
p_{i}=e dr$ with distance $r_{0}$. While for two trapped ions with
a distance $r_{0}$, the interaction by laser induced displacements
$dr$ is $2(\sigma_{z1}^{s}\sigma_{z2}^{s}) e^{2}dr^{2}/r_{0}^{3}$.
The interaction between the charge and the ion is hence a factor
of $r_{0}/2dr$ stronger than the familiar dipole-dipole couplings
encountered in quantum optics, as is shown in Fig.~\ref{fig:2} (a)
and (b).

\subsection{Realization of the Superconducting Qubit}

\label{sec:2.2} 
\begin{figure}[tbp]
\begin{center}
\includegraphics[width=7cm]{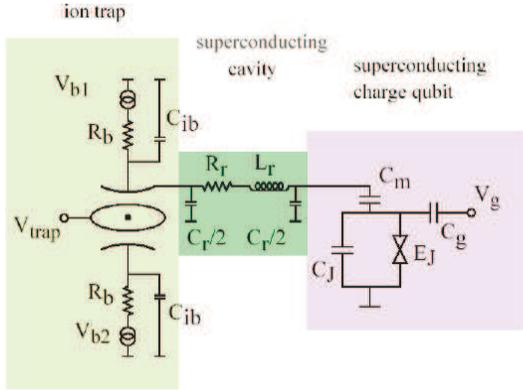}
\end{center}
\caption{ Schematic circuit of the system: ion trap, superconducting cavity
and charge qubit.}
\label{fig:3}
\end{figure}

In our previous paper\cite{interfacing}, we showed
that the hybrid qubit can be formed by connecting an ion with a
superconducting charge qubit, i.e. a superconducting island
connected to a high resistance tunnel junctions (see
Fig.~\ref{fig:3}). In the charge qubit, the quantum
two level system is made of the charge states $|0\rangle =|n\rangle $ and $%
|1\rangle =|n+1\rangle $, with $n$ the number of Cooper pairs on
the island, and the Hamiltonian is $H_{q}$ with $E_{x}=E_{J}$, the
Josephson energy, and $E_{z}=E_{c}(C_{g}V_{g}/2e)$, the charge
bias due to gate voltage $V_{g}$ and charging energy
$E_{c}$\cite{superconducting_qubits}. Other solid-state systems
such as a double quantum dot qubit can be considered within
similar framework. Instead of a direct coupling of the ion to the
charge qubit, we introduce a superconducting cavity between them,
which increases the coupling and provides shielding of the charge
qubit from the stray photons of the trap. The cavity is
characterized by the capacitance $C_{r}$ and the inductance
$L_{r}$ of the cavity, and has eigenfrequencies $\omega
_{n}=n/\sqrt{C_{r}L_{r}}$, with $n$ an integer. In our scheme the
cavity is much shorter than the microwave wavelength, so that the
cavity can be described as two phase variables $\psi _{1,2}$
corresponding to the phases at the ends of the cavity. At one end,
the cavity as part of the trap electrode couples with the ion. At
the other end, the cavity couples with the charge qubit via the
capacitance $ C_{m}$. The Lagrangian of the connected system is
\begin{equation}
\begin{array}{rcl}
\mathcal{L} & = & \displaystyle\frac{C_{r}}{4}\left( \dot{\psi}_{1}^{2}+\dot{%
\psi}_{2}^{2}\right) +\sum_{k=i.i2}\frac{C_{k}}{2}(V_{k}-\dot{\psi}_{1})^{2}-%
\frac{(\psi _{1}-\psi _{2})^{2}}{2L_{r}} \\
& + & \displaystyle\frac{C_{t}}{2}\dot{\varphi}^{2}-C_{g}V_{g}\dot{\varphi}%
-E_{J}\cos{(2e\varphi/\hbar)} +\mathcal{L}_{ion} \\
& + & \displaystyle\frac{C_{m}}{2}(\dot{\psi}_{2}-\dot{\varphi})^{2}+(V_{i}-%
\dot{\psi}_{1})\frac{e\hat{x}}{d_{i}}%
\end{array}
\label{lagrangian_total_s_q}
\end{equation}%
where the first line describes the cavity modes with coupling to the
voltages $V_{i}$ and $V_{2}$ of the trap electrodes via capacitance $C_{i}$
and $C_{i2}$. The second line describes the charge qubit and the ion with $%
C_{t}=C_{J}+C_{g}$. The third line is the capacitive couplings
between the cavity, the ion and the charge qubit, with $d_{i}$ the
distance between the trap electrodes.

After integrating out the cavity modes, the effective coupling is \cite%
{interfacing}%
\begin{equation}\label{Hint}
\displaystyle H_{int}^{(2)}=\frac{e^{2}}{C_{\Sigma }}\frac{C_{m}}{C_{t}}%
\sigma _{z}^{q}\frac{\hat{x}}{d_{i}}  \label{int_i_t_c}
\end{equation}%
where $C_{\Sigma }\approx C_{r}$ when $C_{r}\gg
C_{m},C_{i},C_{t}$. By inserting the cavity, the coupling strength
increases by a factor of $10L/d_{i}$ with $10$ a geometry factor,
as shown in Fig.~\ref{fig:2} . Typical parameters are: cavity
length $L=40\,\mathrm{\mu m}$, $ C_{r}=3\times
10^{-15}\,\mathrm{F}$, $L_{r}=3\times 10^{-13}\,\mathrm{H}$, $
C_{m}=10^{-16}\,\mathrm{F}$, $C_{J}=10^{-16}\,\mathrm{F}\sim
C_{g}$, $ E_{c}=100\,\mathrm{GHz}$,
$e^{2}/2C_{r}=10\,\mathrm{GHz}$, $\omega _{\nu }=1\,\mathrm{MHz}$,
and $d_{i}=20\,\mathrm{\mu m}$. With a laser induced separation of
$\langle \hat{x}\rangle =\pm 200\,\mathrm{nm}$, $
H_{int}^{(2)}=2\pi \times 200\,\mathrm{MHz}$. This interaction
results from electrostatic coupling between the cavity, the ion
and the charge qubit. Hence no resonance condition between the
trap, the cavity and the charge qubit is required and no effort is
needed to tune the various systems to match each other.

\subsection{Two Qubit Gate}

\label{sec:2.3}A controlled phase gate $U=e^{-i\left( \pi
/4\right) \sigma _{z}^{s}\sigma _{z}^{q}}$ can be performed on the
ion and the charge qubit. Three phase gates together with single
qubit gates form the swap gate between the two
qubits\cite{Nielsen_Chuang_book} which exchanges the states of the
qubits and is the key step in interfacing the ion and the charge
qubit. Here it is shown the phase gate does not depend on the
initial state of the motion and operates at nanosecond time scale,
much shorter than the trapping period $\omega _{\nu
}^{-1}$\cite{juanjo_fast_gate}. We define the free evolution as
$U_{0}(t)=\exp {(-i\omega _{\nu }t\hat{a} ^{\dagger }\hat{a})}$,
the entanglement between the internal mode
and the motional mode as $U_{l}(z_{l}n_{l}) = \prod_{m=1}^{n_{1}} \\
\sigma _{x}^{s} \exp {(-iz_{m}\delta k_{l}\sigma
_{z}^{s}\hat{x})}$ achieved by applying laser $\pi $ pulse for
$n_{l}$ times with $z_{m}=(-1)^{m-1}$ and $\delta k_{l}$ the
photon momentum, and the entanglement between the charge qubit and
the motional mode as $U_{2}(\tau _{q})=\exp {(-iH_{int}^{(2)}\tau
_{q}/\hbar)}$ achieved by turning on the capacitive coupling and
decreasing the Josephson energy $E_{J} $.
\begin{figure}[tbp]
\begin{center}
\includegraphics[width=5cm]{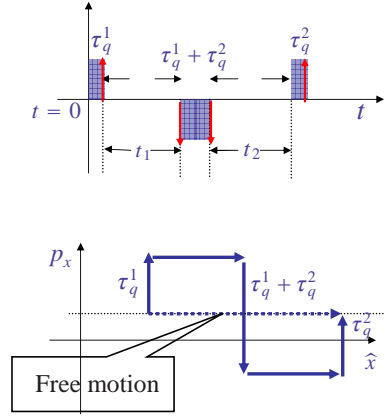}
\end{center}
\caption{The fast quantum phase gate. Top: the pulse sequence of
the gate. Arrow: laser pulse; square: coupling between charge and
ion over a time of  $\tau_q^{1}$, and $\tau_q^{1}+\tau_q^2$ and
$\tau_q^{2}$, and $t_1$ and $t_2$: free evolution. Bottom: the
phase space evolution of the motional mode under the kicks at
qubit states $\sigma_z^q=\sigma_z^s=1$. Dotted line is the free
evolution. } \label{fig:4}
\end{figure}
The gate sequence as is shown in Fig.~\ref{fig:4} contains eight
steps with three laser kicks of $\left\{
n_{l}^{1},-n_{l}^{1}-n_{l}^{2},n_{l}^{2}\right\}$ and three
couplings with charge qubit of durations of $\left\{ \tau
_{q}^{1},-\tau _{q}^{1}-\tau _{q}^{2},\tau _{q}^{2}\right\}$
respectively. These interactions are separated by free evolutions
of durations of $t_1$ and $t_2$. When
$n_{l}^{1}t_{1}=n_{l}^{2}t_{2}$ and $\tau _{q}^{1}t_{1}=\tau
_{q}^{2}t_{2}$, we have
\begin{equation}
\displaystyle U(T)=e^{i\phi ^{\prime
}}U_{0}(T)\exp{\left(-i\frac{e^2 \delta k_{l}\tau
_{q}^{1}n_{l}^{1}t_{1}t_{2}}{2C_r md_{i}^{2}(t_{1}+t_{2})} \sigma
_{z}^{q}\sigma _{z}^{s}\right )},  \label{UT}
\end{equation}
at $\omega _{\nu }t \ll 1$, with a fidelity of $1-O(\omega _{\nu
}^{2}t^{2})$ (Eq.~(\ref{UT}) is exact for free particle). It can
be shown\cite{interfacing} that the speed of the phase gate is
essentially limited by the laser power and the coupling between
the ion and the charge qubit. For the phase gate of $\alpha =\pi
/4$, with $t_{1}=t_{2}=5\,\mathrm{nsec}$, $\delta
k_{l}=10^{8}\,\mathrm{m}^{-1}$, and $n_{l}^{1,2}=10$, the gate
time is $T=14\,\mathrm{nsec}$ for ${^{9}}\mathrm{Be}^{+}$ and $
T=26\,\mathrm{nsec}$ for $^{43}\mathrm{Ca}^{+}$.

\section{Decoherence}
\label{sec:3} In quantum computing, it is important that the
qubits remain in coherent superposition of the quantum states.
Many environmental factors can destroy the coherence of a quantum
system. In the hybrid qubit, noise of both qubits and of the
connecting circuit causes decoherence. In this section, we study
two major sources of decoherence: the charge fluctuations of the
solid-state qubit and the cavity dissipation due to the stray
photons.

\subsection{Charge Fluctuations}

\label{sec:3.1} In solid-state qubits, one major noise source is
charge fluctuations in the substrate or in the gate electrodes due
to the imperfections in fabrication. The charge fluctuations can
be described as dipole jumps in defects which interacts with the
qubit or as charge hoppings near gates which induces image charge
on the gate. Due to interaction with their environments, the
hopping and the dipole jump occur with a $1/f$ dependent spectrum
and cause nonequilibrium and non-Markovian noise to the charge
qubit. It has been shown that the decoherence of the charge qubit
is dominated by this noise with a decoherence time shorter than
nanoseconds when the qubit is away from the degenerate
point\cite{nakamura_1_f}. In the superconducting charge qubit, the
main contribution of the charge noise spectrum is below MHz.

In the hybrid qubit scheme during the storage time, the charge
qubit is static at the degenerate point with a bias voltage
$V_{g}=0$ and $E_{J}\sim 10\,\mathrm{GHz}$:
$H_q=E_{J}\sigma_{z}^{q}/2$. Hence the Josephson energy protects
the charge qubit from charge fluctuations, and the decoherence
time approaches microseconds. However one key step in the
controlled phase gate is the evolution $U_{2}(\tau _{q})$ which
requires that the charging bias is much stronger than the
Josephson energy $\bar{E}_{z}=E_{c}\left(
C_{g}\bar{V}_{g}/2e\right) \gg \bar{E}_{J}$, where $\bar{E}_{J}$
is the reduced Josephson energy during the phase gate with
$\bar{E}_{J}\sim 100\mathrm{MHz}$ and $E_{J}\gg \bar{E}_{J}$. As a
result, the charge qubit is exposed to charge fluctuations in the
environment and subject to strong decoherence during the phase
gate.  The quantum phase gate is performed on a time scale of
$20\,\mathrm{nsec.}$ This indicates that charge noise will induce
serious decoherence to the system.

The charge fluctuations $\delta Q_{g}$ can be expressed as a
voltage noise with $\delta Q_{g}=C_{g}\delta V_{g}$. Adding the
voltage noise to the Hamiltonian, we have: $H_{q}=E_{c}\left(
C_{g}\bar{V}_{g}+C_{g}\delta V_{g}\right)/2e +
\bar{E}_{J}\sigma_{z}^{q}/2$ where the dynamics of the $
\bar{E}_{J}$ term, being reduced during the gate, will be
neglected in the following discussion. The noise $\delta V_{g}$ is
a stochastic operator with a spectrum $\delta V_{\omega }^{2}=2\pi
\langle \delta V\left( \omega \right) \delta V\left( \omega
^{\prime }\right) \rangle \delta \left( \omega +\omega ^{\prime
}\right) $ and $\delta V\left( \omega \right) =\int dt^{\prime
}e^{i\omega t^{\prime }}\delta V\left( t\right) $, determined by
the environment. The low frequency noise has $\delta V\left(
\omega \right) \propto 1/\omega$. The gate transformation in
Eq.~(\ref{UT}) is now
\begin{equation}
\displaystyle U(T)=e^{i\phi ^{\prime
}}U_{0}(T)e^{-i\bar{E}_{z}T^{\prime }/\hbar }e^{-i\sigma _{z}\phi
_{v}\left( t\right) }e^{-i\alpha \sigma _{z}^{q}\sigma _{z}^{s}}
\label{UT2}
\end{equation}%
where two phase factors are added compared with Eq.(\ref{UT}): the
dynamical phase $e^{-i\bar{E}_{z}T^{\prime }}$ due to the voltage
bias in the Hamiltonian; and the stochastic phase $e^{-i\phi
_{v}\left( t\right) }$ due to the charge noise. We have $\phi
_{v}\left( t\right) =$ $\left( E_{c}C_{g}/2e\hbar \right)
\int_{0}^{t}dt^{\prime }$\ $\delta V_{g}\left( t^{\prime }\right)
$. The dynamical phase does not affect the gate, but the
stochastic phase causes serious decoherence.

In the following we extend the quantum phase gate in
\cite{interfacing} and present a scheme that overcomes the effect
of the low frequency noise. Instead of directly applying the eight
steps of the phase gate, we divide the gate into shorter pieces to
improve its resistance to the charge noise. Let $\tau =T/N$ be the
unit of improved gate, with $N \gg 1$ being an even integer. The
gate consists of $N$ pieces each contributing a phase $\alpha
=\pi/4N$ to the quantum phase gate. The gate sequence is now
$U(T)=(\sigma _{x}^{q}U(\tau ))^{N}$, where after each interval
$\tau$, a $\pi$ pulse is applied to the charge qubit to flip the
charge state. We assume $\tau $ is of the order of or below
nanoseconds. The $\pi$ pulses can be achieved by increasing the
effective Josephson energy $\bar{E}_J$ to about $10\,\textrm{GHz}$
for short intervals of subnanoseconds; the fidelity of the $\pi$
pulses is mainly limited by the switching time of the Josephson
energy and hence the switching time of the flux in the qubit
circuit.

The gate evolution is
\begin{equation}
\displaystyle U\left( t\right) =e^{i\phi ^{\prime }}e^{i\sigma _{z}\bar{\phi}%
\left( t\right) }U_{0}(T)e^{-i\frac{\pi }{4}\sigma _{z}^{q}\sigma
_{z}^{s}}
\end{equation}%
where instead of the phase $\phi _{v}\left( t\right)$, the random
phase becomes $\bar{\phi}\left( t\right) =\left(
E_{c}C_{g}/2e\right) \int dt^{\prime }\delta V_{g}\left( t^{\prime
}\right)\, g\left( t\right )$ with
\[
\displaystyle g\left( t\right) = \{ \begin{array}{clc} 1, & t\in
\left[ 2n\tau ,2n\tau +\tau \right), & \\  -1, &  t\in \left[
2n\tau +\tau ,\left( 2n+1\right) \tau \right). & \\
\end{array}
\] which is created by the periodic charge flips.
Mathematically, the function $g\left( t\right)$ can be decomposed
into triangular functions: $g\left( t\right) =\sum_{m}\left( 4/\pi
m\right) \sin \omega _{m}t$ with frequencies $\omega _{m}=m\left(
2\pi /\tau \right)$ -- multiples of $\omega_{1}=2\pi /\tau$. This
procedure is equivalent to shifting the noise spectral density of
the charge fluctuations by frequencies $\omega _{m}$. In the
lowest order, the decoherence rate can be calculated by
\[
\gamma_{v}=\frac{\partial \langle e^{i\bar{\phi}( t)}\rangle
}{\partial t} \quad \mathrm{and} \quad \langle e^{i\bar{\phi}( t)
}\rangle =e^{-\frac{\langle \bar{\phi}^{2}(t)\rangle}{2}}.
\] We have
\begin{equation}
\begin{array}{c}
\displaystyle \left\langle \bar{\phi}^{2}\left( t\right)
\right\rangle
_{ev}=\left( \frac{E_{c}C_{g}}{2e}\right) ^{2}\frac{4}{\pi ^{2}}\sum_{n,m}%
\frac{1}{mn}\cdot \\
\displaystyle \int \frac{d\omega }{2\pi }\delta V_{\omega
}^{2}\int_{0}^{t}dt^{\prime }e^{it^{\prime }\left( \omega
_{n}-\omega \right) }\int_{0}^{t}dt^{\prime \prime }e^{-it^{\prime
\prime }\left( \omega
_{m}-\omega \right) }.%
\end{array}
\label{phi2}
\end{equation}%
Consider $\omega _{1}t\gg 1$. The variance of the random phase is
\begin{eqnarray*}
\displaystyle \left\langle \bar{\phi}\left( t\right)
^{2}\right\rangle _{ev}
&=&\displaystyle \left( \frac{E_{c}C_{g}}{2e}\right) ^{2}\frac{2}{\pi ^{2}}%
\sum_{n}\frac{\delta V_{\omega _{n}}^{2}}{n^{2}}t \\
&\leq &\displaystyle \left( \frac{E_{c}C_{g}}{2e}\right) ^{2}\frac{1}{3}%
\delta V_{\max }^{2}t
\end{eqnarray*}%
where in the inequality relation, we replace the spectral density
by the maximal spectral density above $\omega _{1}$
($\omega_{1}\ge \mathrm{GHz}$), and applying the relation:
$\sum_{1}^{\infty }1/n^{2}\newline =\pi ^{2}/6$. This shows that
the decoherence rate is now dominated by noise spectral density
above GHz: $\gamma _{v}\leq \left( E_{c}C_{g}/2e\right) ^{2}\left(
1/3\right) \delta V_{\max }^{2}$, and effect of the low frequency
noise is reduced by the charge flips. Note at GHz frequencies, the
charge noise is mainly thermal noise of the connecting circuits,
e.g. Johnson-Nyquist noise of the resistances in the circuit. In
experiments, this noise induces a decoherence rate slower than
MHz\cite{Saclay_exp}. By flipping the charge qubit at very short
intervals $m\tau $, a spin-echo type of spectral modulation is
achieved which engineers the noise spectral density and hence
protects the charge qubit from the low frequency noise.

In a scalable scheme, to avoid affecting other charge qubits by
the fast flips of one qubit, we consider using local
superconducting wire to control the flux in the double junction of
the charge qubit\cite{superconducting_qubits}. Because the flux
generated by a wire decreases with the distance to the wire $r$ as
$1/r$, only nearby charge qubits will be affected by this flux.
Meanwhile, we design the charge biases $E_c(C_gV_g/2e)$ of the
qubits in a neighborhood to have differences above
$5\,\mathrm{GHz}$.  With a flipping rate of GHz, the off resonance
in the other qubits will prevent their flipping.

\subsection{Cavity Dissipation}

\label{sec:3.2} Another source of decoherence we consider is the
losses in the superconducting cavity which introduces decoherence
to the qubits. At low temperature in a superconductor, the
quasiparticle density decreases exponentially with temperature:
$n_{n}=n_{0}\exp {(-2\Delta /k_{B}T)}$, so that the dissipation
due to quasiparticle conduction can be neglected. However, when
laser photons, e.g. from the laser driven ion, are scattered to
the superconductor, quasiparticles are excited and dissipation
increases.  In our previous paper\cite{interfacing}, we estimated
the effect of the induced quasiparticles on decoherence. Here, we
present a path integral approach to calculate the decoherence
rate.

We model the dissipation of the cavity as a resistor in series
with the cavity inductance. The resistance is
$R_{r}=R_{n}(n_{ex}/n_{0})$\cite{jackson_em,tinkham_superconductivity},
where $R_{n}$ is the normal state resistance of the
superconductor. Considering a  laser power of mW, and assume the
stayed photons consist $10^{-6}$ of the laser power. In a duration
of $100\,\mathrm{nsec}$, the photons excite quasiparticles that
can be modeled as a resistance of $R_{r}=R_{n}/10^{5}$.

The dissipation can be calculated with the standard
Caldeira-Leggett formalism\cite{grabert_phys_rep} which treats the
environments as an oscillator bath that couples with the qubit
linearly. We derive the effective spectral density of the cavity
resistance. The Hamiltonian including the bath and the cavity
modes is
\begin{equation}
\displaystyle H_{R}=H_{cav}+\tilde{\psi}_{k}\sum_{j}\lambda
_{j}x_{j}+\sum_{j}\left( \frac{p_{j}^{2}}{2m_{j}}+\frac{m_{j}\omega _{j}^{2}%
}{2}x_{j}^{2}\right)  \label{c_t_bath}
\end{equation}%
where $H_{cav}$ is the cavity Hamiltonian derived from
Eq.(\ref{lagrangian_total_s_q}); $x_{j}$s are the coordinates of
the oscillator modes in the bath, $p_{j}$s are the momenta of the
oscillator modes, $m_j$s the mass, $\omega_{j}$s the oscillator
frequencies, with the last term being the Hamiltonian of the
bosonic modes. The couplings between the oscillators and the
cavity mode $\tilde{\psi}_{k}$ are the $\lambda _{j}$s which
determine the noise spectral density $J_{0}(\omega)=\sum \lambda
_{j}^{2}/2m_{j}\omega _{j}\delta (\omega -\omega _{j})$. In the
case of a resistance, we have $J_{0}(\omega)=\omega R_{r}^{-1}$.
Note the complete Hamiltonian is $H_{t}+H_{R}$ including the ion,
the charge qubit, the cavity and the bath. In the path integral
approach, the harmonic oscillator degrees of freedom can be
integrated exactly as we are only concerned with the dynamics and
decoherence of the qubits. In the following we take the charge
qubit as an example to study the decoherence. The same approach
can be applied to the motion of the ion. After integrating out
both the cavity and the bath modes, the effective action of the
charge qubit is
\begin{equation}
\begin{array}{c}
\displaystyle S_{eff}^{E}[\bar{\varphi}]=\int_{0}^{\hbar \beta }d\tau \frac{%
C_{t}}{2}\dot{\varphi}^{2}-E_{J}\cos{( 2e\varphi/\hbar)} \\
\displaystyle +\frac{1}{2}\int_{0}^{\hbar \beta }d\tau
\int_{0}^{\hbar \beta
}d\sigma k(\tau -\sigma )\dot{\varphi}(\tau )\dot{\varphi}%
(\sigma )%
\end{array}
\label{effective_action_3}
\end{equation}%
where $\varphi$ is the gauge invariance phase of the charge
qubit\cite{grabert_phys_rep}, and $\beta=1/k_BT$ with $T$ being
the temperature. The function $k$ is
\begin{equation}
\displaystyle k(\tau )=-\frac{(C_{m}/2C_{t})^{2}}{\hbar \beta
(C_{r}+C_{m})}\sum_{n=-\infty }^{+\infty }\frac{\nu
_{n}^{2}e^{i\nu _{n}\tau }}{(\nu _{n}^{2}+\omega _{r}^{2}+|\nu
_{n}|\hat{\gamma}(|\nu _{n}|))}, \label{kbar_tau}
\end{equation}%
with $\omega _{r}=2/\sqrt{(C_{r}+C_{m})L_{r}}$ and $\nu _{n}=2\pi
n/\hbar \beta $ being the Matsubara frequencies at integer $n$.
Let the Fourier transformation of $k(\tau )$ be
$\widetilde{k}(i\nu _{n})=\int_{0}^{\hbar \beta }d\tau k(\tau
)\exp {(i\nu _{n}\tau )}$. The retarded noise spectral function is
\begin{equation}
\displaystyle \widetilde{k}(i\nu _{n}=\omega +i\delta
)=(\frac{C_{m}}{2C_{t}})^{2}i\omega Z_{\mathrm{eff}}(\omega )
\label{kbar_retarded}
\end{equation}%
where the spectral density is characterized by an effective
impedance $Z_{\mathrm{eff}}$ described as a capacitor
$(C_{r}+C_{m})/4$ in parallel to the series of the inductor
$L_{r}$ and the resistor $R_{r}$. When $ \omega \ll
1/\sqrt{L_{r}C_{r}}$, we have $Z_{\mathrm{eff}}\approx R_{r}$.

The action in Eq.~(\ref{effective_action_3}) describes the charge
qubit interacting with a fluctuating field with a spectral density
\begin{equation}\label{Jomega}\begin{array}{c}
\displaystyle J_{\mathrm{eff}}\left( \omega \right)
=\mathrm{Im}[\widetilde{k}(\omega +i\delta ) ]
\\ [2mm] \displaystyle =(\frac{C_{m}}{2C_{t}})^{2}\omega Z_{\mathrm{eff}}(\omega )\coth
\left( \frac{\hbar \omega }{2k_{B}T}\right) \end{array}
\end{equation}
which is derived from the above discussion. The decoherence rate
of the charge qubit $\gamma _{r}^{q}$ can be derived from
$J_{\mathrm{eff}}\left( \omega \rightarrow 0\right) $ according to
the fluctuation dissipation theorem:
\begin{equation}
\displaystyle \gamma _{r}^{q}\approx
\frac{R_{r}}{R_{k}}\frac{2k_{B}T}{\hbar
}(\frac{C_{m}}{2C_{t}})^{2}
\label{decoherence_charge_motion}
\end{equation}
where $R_{k}=\hbar /(2e)^{2}$ is the quantum resistance. At a
temperature of $T=100\, \mathrm{mK}$, we have $\gamma
_{r}^{q}=50\,\mathrm{msec}^{-1}$. This shows that the dominant
decoherence is not due to the cavity
loss\cite{superconducting_qubits,ion_trap_exp_review_wineland},
but most likely due to the charge noise.

\section{Experimental Issues}

\label{sec:4}  Combining systems as different from each other as
the ion trap and the solid-state qubit is a challenge for existing
experimental techniques. Questions arise such as whether the two
systems are compatible and whether the techniques developed for a
conventional system can be applied to the combined system. In this
section we investigate several experimental issues of the combined
system, including the fast switch during the swap gate, the
balance circuit that decouples the charge qubit from the ac
driving of the trap and the scalability issue.

\subsection{Fast Switch}

\label{sec:4.1} In the quantum phase gate, the laser pulse, the
coupling between the ion and the charge, and the free evolution
are applied alternatively, which requires a fast switch that can
turn on and turn off the capacitive coupling in a time scale
shorter than the gate time\cite{interfacing}. Various switch
circuits have been studied with mesoscopic electronics such as the
superconducting field effect transistor, superconducting single
electron transistor, and the $\pi$-junctions of high T$_{c}$
materials\cite{superconducting_qubits}. In this section we study a
fast electronic switch made of dc SQUID.

\begin{figure}[tbp]
\begin{center}
\includegraphics[width=7.5cm]{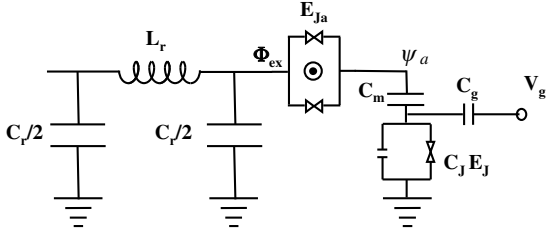}
\end{center}
\caption{ Schematic circuit of superconducting dc SQUID switch
inserted between the cavity and the charge qubit. The island
between the switch and the capacitor $C_{m}$ is labeled
$\psi_{a}$.} \label{fig:5}
\end{figure}
The switch, inserted between the superconducting cavity and the
capacitor $C_{m}$ is shown in Fig.~\ref{fig:5}. The switch is made
of two large Josephson junctions with a Josephson energy $E_{Ja}$
much larger than the Josephson energy $E_{J}$ of the charge qubit
forming a dc SQUID, and the charging energy of the junctions is
negligible. In the dc SQUID geometry, the switch is described as a
junction with an effective Josephson energy
$E_{a}=2E_{Ja}\cos{(\mathrm{\pi\Phi_{ex}/\Phi_{0}})}$ depending on
the external flux $\mathrm{\Phi_{ex}}$ in the SQUID loop, with
$\mathrm{\Phi_{0}}$ the flux quantum. This shows that when
$\mathrm{\Phi_{ex}=\Phi_0/2}$, the connection between the cavity
and the charge qubit is cut off by the SQUID. Below, we derive the
coupling between the charge qubit and the ion in the presence of
the switch.

We introduce four phase variables to describe the circuit in
Fig.~\ref{fig:5}: $\psi_{1,2}$ of the left and the right ends of
the cavity, $\psi_{a}$ of the island between the switch and the
capacitor $C_{m}$, and $\varphi$ of the charge qubit. The
Lagrangian of the system can be derived by replacing the term
$C_{m}(\dot{\psi}_2-\dot{\varphi})^2/2$ in
Eq.~(\ref{lagrangian_total_s_q}) with
\begin{equation}\label{switch_L}
\displaystyle  \frac{C_{m}}{2}(\dot{\psi}_a-\dot{\varphi})^2 +
E_{a} \cos{(\frac{2e(\psi_{2}-\psi_{a})}{\hbar})}
\end{equation}
which includes the capacitive energy of $C_{m}$ and the Josephson
energy of the SQUID. The Hamiltonian can be derived as:
\begin{equation}\label{switch_H}\begin{array}{ccl}\displaystyle
H_{sw} &=&\displaystyle
\frac{p_1^2+p_2^2}{C_r}+\frac{2p_1e}{C_r}\frac{\hat{x}}{d_i} +
\frac{(\psi_1-\psi_2)^2}{2L_{r}} + \frac{p_a^2}{2C_a} \\
[2mm] & \displaystyle + & \displaystyle
\frac{(p_\varphi+C_gV_g)^2}{2C_t}-E_J\cos{(2e\varphi/\hbar)} \\
[2mm] &\displaystyle +& \displaystyle
\frac{p_a(p_\varphi+C_gV_g)}{C_t} -E_{a}
\cos{(\frac{2e(\psi_{2}-\psi_{a})}{\hbar})}
\end{array}
\end{equation}
where $p_{i},\,i=1,2,a,\varphi$ are the conjugate variable of the
corresponding phase variables and $C_{a}=C_m C_t/(C_m+C_t)$. Here
the charging energy of the island between the switch and $C_m$ is
$p_{a}^{2}/2C_{a}$, and the last two terms are the coupling
between $p_a$ and the cavity, and the coupling between $\psi_{a}$
coupling and the charge qubit. When
$\cos{(\mathrm{\pi\Phi_{ex}/\Phi_{0}})}=0$, i.e.
$\mathrm{\Phi_{ex}=\Phi_0/2}$, the last term in
Eq.~(\ref{switch_H}) disappears, and hence the cavity together
with the ion is disconnected from the island $\psi_{a}$ and the
charge qubit. After integrating out $\psi_{a}$, the charge qubit
Hamiltonian is
$(p_\varphi+C_gV_g)^2/2(C_m+C_t)-E_J\cos{(2e\varphi/\hbar)}$
consistent with the Hamiltonian without the switch.  This shows
that the coupling between the charge qubit and the ion can be
turned off by applying a $\pi$-flux in the SQUID loop.

Now we calculate the effective coupling between the charge qubit
and the ion with the switch on. The large Josephson junction can
be modeled as an inductance with
$L_{\mathrm{eff}}=(\hbar/2e)^2/E_{a}$ and the energy is
$(\psi_{2}-\psi_{a})^2/2L_{\mathrm{eff}}$. The quadratic
Hamiltonian of $\psi_{1,2,a}$ and $p_{1,2,a}$ can be diagonalized
into secular modes $\psi_k=\sum v_{ki}\psi_{i}$ and $p_k=\sum
v_{ki}^\star p_i$ with $i=1,2,a$ and $k=A,B,C$. One of these modes
$\psi_C$ is $\psi_C=\sum \psi_i/\sqrt{3}$ with $p_C=\sum p_i
/\sqrt{3}$ and the secular value zero. Fig.~\ref{fig:5} shows that
$p_C=0$. Hence we derive:
\begin{equation}\label{switch_H2}\begin{array}{ccl}
\displaystyle H &\displaystyle =& \displaystyle\sum_{k=A,B}
\frac{p_k^2}{C_r}+\frac{\psi_k^2}{2L_k} +
\frac{(p_\varphi+C_gV_g)^2}{2C_t}-E_J\cos{(2e\varphi/\hbar)}
\\ [3mm] &\displaystyle +& \displaystyle \sum_{k=A,B} 2p_k \left
(\frac{e\hat{x}}{d_i}v_{k1}+\frac{\sqrt{C_rC_a}(p_\varphi+C_gV_g)}{\sqrt{2C_t}}v_{k3}
\right )
\end{array}
\end{equation}
where $L_{A,B}$ are functions of $L_{r}$ and $L_{\mathrm{eff}}$
and are the other two secular values besides zero. Integrating
$\psi_{A,B}$, the effective interaction between the charge qubit
and the ion is $H_{int}^{sw}=(e^2C_m\hat{x}/C_rC_td_i)\sigma_z^q$,
exactly the same as that in Eq.~(\ref{Hint}). At the same time,
the integration also gives a correction to the charge qubit which
recovers the form of $(p_\varphi+C_gV_g)^2/2(C_m+C_t)$. This shows
that by inserting a switch with
$\hbar/\sqrt{L_{\mathrm{eff}}C_r}\gg E_J$, the coupling strength
is not affected; while when the critical current disappears, the
coupling disappears as well.

The performance of switch is limited by the speed of switching the
flux $\Phi_{ex}$ in the SQUID and by the incomplete turning off of
the switch. This may be overcome by inserting a $\pi$-junction
into the circuit instead of applying magnetic
flux\cite{superconducting_qubits}. In practice, the large
junctions of the SQUID couple with flux noise or current noise,
and may bring decoherence to the circuit. The decoherence of
various switches was studied in \cite{storcz_wilhelm}.

\subsection{The Balance Circuit}

\label{sec:4.2} In standard ion traps, electromagnetic fields are
applied to achieve trapping of the ions. For example, in a Paul
trap, typically an ac voltage of 100 -- 250 MHz and 30 -- 50 Volts
is applied on the electrodes; in a Penning trap, a magnetic field
gradient is applied.  The coupling between the ion and the charge
qubit not only brings interaction between the two qubits, but also
connects the driving fields with the charge qubit. However, the
superconducting charge qubit can not coexist with strong external
fields. Here we show that the external field can be canceled by
designing a balance circuit.

\begin{figure}[tbp]
\begin{center}
\includegraphics[width=3.5cm]{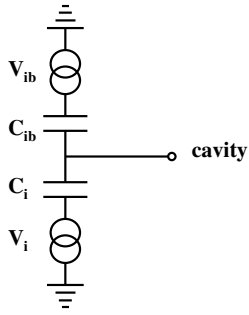}
\end{center}
\caption{The balance circuit. The ac voltages $V_{i}$ and $V_{ib}$
are applied to the trap electrodes. This circuit is connected to
the superconducting cavity. } \label{fig:6}
\end{figure}
We consider a Paul trap connecting with the superconducting
cavity. As is shown in Fig.~\ref{fig:6} and described by the $\sum
C_k(V_k-\dot{\psi_1})^2/2$ term in
Eq.~(\ref{lagrangian_total_s_q}), the cavity is part of the
electrodes and is coupled to the driving voltages via the
capacitors $C_k$. These couplings contribute to the Hamiltonian as
\begin{equation}\label{Hv}
\displaystyle \frac{e^{2}}{C_{\Sigma }}\frac{C_{m}}{C_{t}}%
\frac{(C_iV_i+C_{ib}V_{ib})}{e}\sigma_{z}^{q}
\end{equation}
and can have significant influence on the charge qubit. However,
by choosing the balance condition $C_iV_i+C_{ib}V_{ib}=0$, this
coupling disappears, and the charge qubit is protected in an
experiment. The balance condition discussed above can only be
achieved approximately due to the inaccuracy in the controlling of
the voltages. With an inaccuracy of $10^{-4}\,\mathrm{V}$ in the
voltage sources, and $C_i\sim C_t/10$ ($d_i=20\,\mathrm{\mu m}$
and area of electrodes $10^{-12}\,\mathrm{\mu m}^2$), the coupling
of the voltage sources with the charge qubit $\delta E \sigma_z^q$
is $\delta E=100\,\mathrm{MHz}$, much less than the charging
energy $e^2/C_r$ and the coupling in Eq.~(\ref{Hint}). In
addition, the frequencies of the driving voltages is around 100 --
200 MHz, much less than the charge qubit energy; hence this
inaccuracy and moreover the coupling between the charge qubit and
the driving voltages does not bring serious effect on the charge
qubit.

\subsection{Scalability}
\begin{figure}[tbp]
\begin{center}
\includegraphics[width=8.0cm]{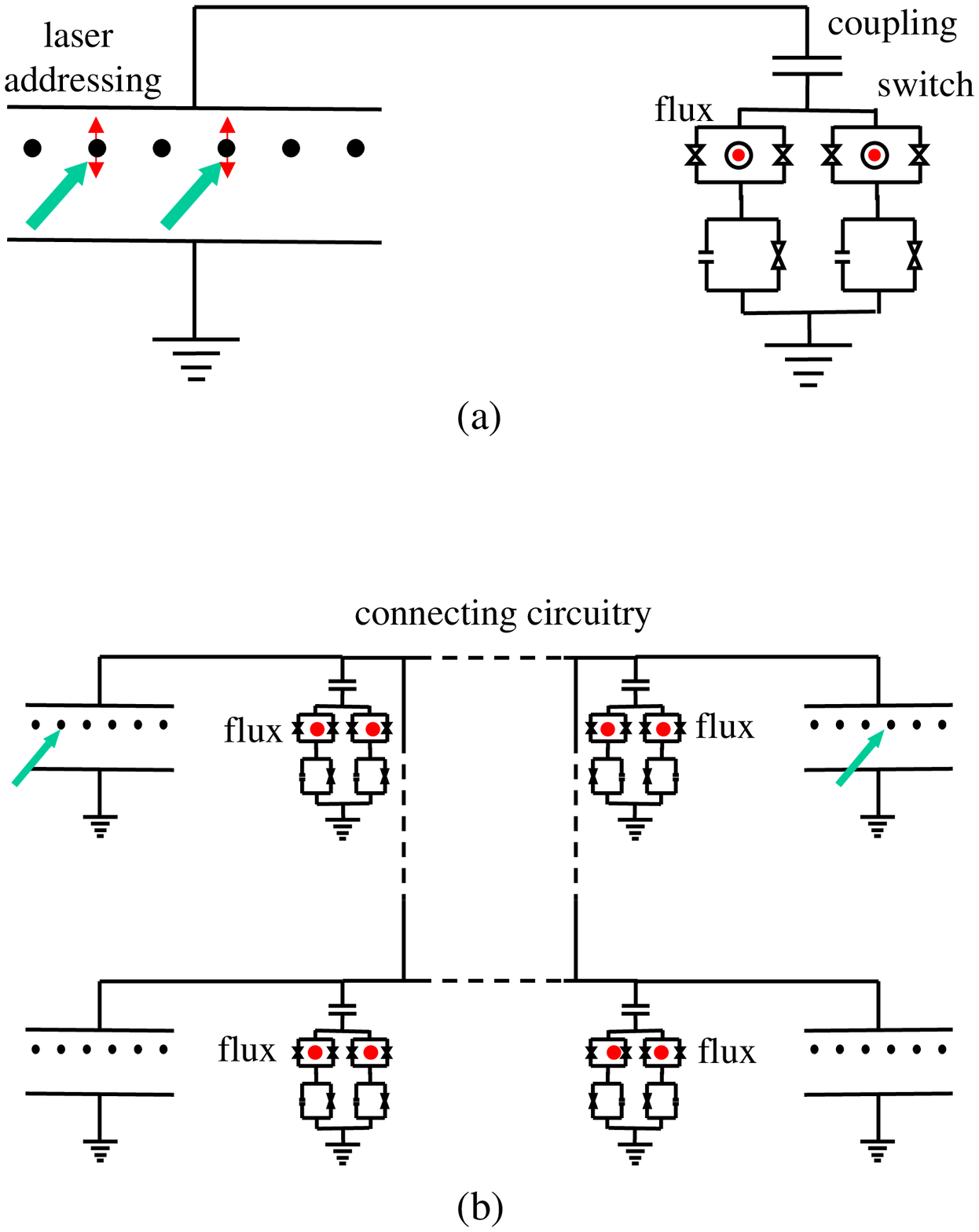}
\end{center}
\caption{Top: an array of ions in a linear trap couple with the
charge qubits via the cavity. Bottom:  four hybrid systems couple
with each other via superconducting transmission line or
capacitive coupling (dash lines represent the coupling circuit). }
\label{fig:7}
\end{figure}

\label{sec:4.3} With the hybrid system, we hope to achieve
scalable quantum computing without moving ions. To exploit the
advantages of both the ion and the charge qubit, we consider the
scheme shown in Fig.~\ref{fig:7} as an example. Compared with pure
ion trap qubits, it is harder to fabricate large numbers of the
hybrid qubits which consist of an ion and a charge qubit each.
Instead, we design the hybrid system in the form of clusters: an
array of ions trapped in a linear trap are connected with two
charge qubits via a superconducting cavity -- a cluster of small
numbers of ion qubits. The charge qubits capacitively couple with
the cavity, and the couplings are controlled by the switches made
of the SQUIDs.

During a controlled phase gate between ions in the same cluster, 
one ion is addressed by a polarization dependent laser pulse that 
pushes the ion in the transverse direction which is shown by the 
arrows in  Fig.~\ref{fig:7}. All the other ions in the array are 
unaffected. This is followed by turning on the switch of one charge 
qubit to allow the coupling between the charge qubit and the 
addressed ion. After the gate sequence in Eq.~(\ref{UT}), the 
phase gate is obtained between this ion and this charge qubit, and 
subsequenctly the swap gate. For the other ions, the evolution is 
$U^\prime(T)=\exp{(-i\omega_\nu t\hat{a}^\dagger 
\hat{a}+i\phi^{\prime\prime})}$ which is nothing but free 
evolution; hence the other ions are exempted from the controlled 
gate. The same procedure is then applied to the other ion 
involved in the controlled gate after which the two qubit gate is 
performed on the charge qubits. Note when addressing an ion in 
its transverse direction, the Coulomb interaction between the 
ions contributes a small force in the same direction of the 
transverse motion.  This force is smaller than the force of the 
trapping potential when the distance between the ions is longer 
than microns at a trapping frequency of $10\,\mathrm{MHz}$.

Based on this scheme, multiple clusters of the hybrid system can
be fabricated. In Fig.~\ref{fig:7}, we show four such systems each
including an array of ion in the ion trap and two charge qubits.
The clusters are coupled by capacitively connecting a charge qubit
in one cluster with another charge qubit in another clusters. With
superconducting transmission lines, distant qubits with a
separation of centimeters can be connected with an interaction
strength of $100\,\mathrm{MHz}$\cite{Girvin_Schoelkopf}. Quantum logic gates on qubits in
the same cluster can be performed on the charge qubits in the same
cluster. Quantum logic gates on qubits in distant clusters can be
performed via the capacitive connections.  Note when there are
many clusters, connections between neighboring clusters are
sufficient to obtain interactions between qubits in any two
clusters.

\section{Discussions and Conclusion}

\label{sec:5} We studied several issues in the interfacing of ion
trap qubits and solid-state charge qubits: decoherence, coupling
mechanism, switching of coupling, and scalability. We calculated
the decoherence due to the charge noise and the dissipation of the
cavity, and presented a gate scheme that can overcome the charge
noise. We analyzed a crucial element of this scheme: a fast switch
that is made of a dc SQUID and can turn off the coupling between
the charge qubit and the ion. We also present an example of
scalable hybrid schemes where the ion qubits are aligned in
clusters -- an array of ions coupling with charge qubits.  It is
shown from these discussions that the hybrid system is a scalable
quantum computing system that may be able to exploit and combine
the merits of very different systems.  Note we concentrate on a
special example of ion trap qubit coupling with superconducting
systems. Study on interfacing other systems, e.g. nanomechanical
resonator and superconducting
qubit\cite{cpb_mechanical_resonator_schwab}, has been studied.

On the other hand, connecting systems as different as the ion trap
and the charge qubit is very challenging. For example, the charge
qubits have to work at millikelvin temperature in a dilution
fridge. This requires that the ions have to be positioned in the
fridge as well. This also brings up the questions of including the
laser in the fridge. We analyzed several experimental issues in
this paper, such as the balance circuit, the state flipping scheme
and the fast switch. Such issues are technically demanding.  We
would like to point out that the study of hybrid systems is still
at the very beginning and some part of the theory is still
speculative. While we expect more interest and study on
interfacing different systems in near future.

Acknowledgments: Work supported by the Austrian Science Foundation, European
Networks and the Institute for Quantum Information.

\end{document}